lica.it
\documentclass[aps,pra,twocolumn,showpacs]{revtex4}
\usepackage{dcolumn}
\usepackage{bm}
\usepackage{graphicx}
\usepackage{amsmath}
\usepackage{latexsym}
\usepackage{amsfonts}
\usepackage{amssymb}
\usepackage{array}
\usepackage{epsfig}
\usepackage{epstopdf} 
\usepackage{times}

\newcommand{\ket}[1]{\vert#1\rangle}

\newcommand{\proj}[3]{\left\vert#1\rangle_{#2}\langle#3\right\vert}

\newcommand{\nbar}{\overline{n}}

\begin{document}
\title{Passing quantum correlations to qubits using any two-mode state}
\author{Mauro Paternostro$^1$}
\author{Gerardo Adesso$^2$}
\author{Steve Campbell$^1$}
\affiliation{$^1$School of Mathematics and Physics, Queen's University, Belfast BT7 1NN, United Kingdom\\
$^2$School of Mathematical Sciences, University of Nottingham, University Park, Nottingham NG7 2RD, United Kingdom}

\begin{abstract}
We draw an explicit connection between the statistical properties of an entangled two-mode continuous variable (CV) resource and the amount of entanglement that can be dynamically transferred to a pair of non-interacting two-level systems. 
More specifically, we rigorously reformulate entanglement transfer process by making use of covariance matrix formalism.  When the resource state is Gaussian, our method makes the approach to the transfer of quantum correlations much more flexible than in previously considered schemes and allows the straightforward inclusion of the effects of noise affecting the CV system. Moreover, the proposed method reveals that the use of de-Gaussified two-mode states is almost never advantageous for transferring entanglement with respect to the full Gaussian picture, despite the entanglement in the non-Gaussian resource can be much larger than in its Gaussian counterpart. We can thus conclude that the entanglement-transfer map  overthrows the ``ordering" relations  valid at the level of CV resource states. \end{abstract}
\date{\today}

\maketitle

The establishment of a long-haul entangled channel is one of the most important and challenging aims of that part of quantum technology concerned with communication and distributed computation~\cite{huelgacirac}. The amount of experimental and theoretical efforts that have been put into the design and realization of a device allowing such a result with a sufficient degree of reliability is tremendous~\cite{efforts}. Among the rich plethora of schemes put forward so far, an intellectually stimulating and pragmatically inspiring proposal relies on the possibility to quantum mechanically correlate two remote systems by {\it transferring} some pre-available entanglement initially carried by a resource~\cite{ET,altri1,altri2}. This stems as a rather natural way of achieving a distributed channel for quantum communication, especially when the entangled resource is embodied by the continuous variable (CV) state of a photonic system, which has been proven to be a reliable courier of quantum correlations. In fact, such an idea has been implemented, recently, in a setup involving cold atomic ensembles and light fields~\cite{kimble}.

Despite the scheme is attracting an increasing attention also in virtue of its versatile nature, which makes it suitable for implementations in various settings, including cavity and circuit quantum-electrodynamics, the formalism itself is largely in need of development. In its basic form, the protocol requires the arrangement of a bilocal interaction between each remote qubit and one of the modes of the CV photonic resource. The aim is to arrange a situation suitable to the {\it pouring} of quantum correlations from the CV resource to the non-interacting two-level systems in such a way that, upon tracing out the two modes, we are left with an entangled state of the qubits. In this procedure, it is often the case that the explicit form of the CV resource, decomposed in some basis, is needed. This makes the computational efforts required for the evaluation of the degree of transferred entanglement  quite demanding. And yet, it is rather sensible to believe that asking for the full CV state decomposition is probably not at all necessary, especially if one considers the class of Gaussian entangled two-mode states, which are fully determined by assigning only a few of their  statistical properties. This is exactly the main point of the present investigation: we develop a self-consistent formal approach that is able to reconstruct the entries of the two-qubit density matrix resulting from the entanglement transfer process simply by means of statistical properties of the pre-built resource. 
This represents the first step of a theoretical investigation directed towards the generalization and extension of the {\it panorama} of situations where entanglement transfer can be conveniently and almost effortlessly studied. This includes also the experimentally important case of non-Gaussian CV resources having the form of $s$-photon subtracted two mode states~\cite{kitagawa}. The proposed formalism allows us to quantitatively study the performance of this class of states in relation to the tasks discussed here.

The remainder of this paper is organized as follows. In Sec.~\ref{prelim} we lay down the notation and recall some basic definitions and tools useful for the treatment of CV systems. In Sec.~\ref{gaussiano} we attack the case of entanglement transfer from a  resource prepared in a general Gaussian state. We give explicit expressions for the elements of the two-qubit reduced density matrix resulting from the transfer process as functions of the statistical moments of the resource.  We highlight their theoretical and computational convenience by addressing a general two-mode squeezed thermal state, which allows us to unveil the spoiling effect, on the degree of transferred entanglement,  of the resource's thermal character. A similar study is then performed with respect to a dissipation-affected two-mode squeezed state: in both cases we demonstrate that the degree of {\it poured} entanglement is set once the resource is assigned. Sec.~\ref{nongaussiano} tackles the case of experimentally relevant non-Gaussian states obtained by subtracting $s$ photons symmetrically from a two-mode squeezed state~\cite{kitagawa,furusawa}. A series of results are achieved in this case: first, we demonstrate that the Gaussian core of the photon-subtracted resource ({\it i.e.} its statistical moments up to the second) completely specifies the amount of transferred entanglement. However, the use of such a non-Gaussian state is not always advantageous with respect to its Gaussian counterpart (obtained by not subtracting photons): despite photon-subtraction works as an entanglement distillation protocol for the CV state, a two-mode Gaussian state is almost always a more efficient resource for entanglement transfer. We also draw some conjectures on the relation between the discrepancy in the performances of Gaussian-associated and original non-Gaussian states and the degree of non-Gaussianity of a resource~\cite{paris}. In Sec.~\ref{conclusioni} we summarize our findings and highlight the procedure for adapting our technique to the recently proposed protocol for {entanglement reciprocation}~\cite{reciproco}. Finally, in Appendices I and II we provide a few technical steps needed in order to fully understand our formal approach.

\section{Preliminaries}
\label{prelim}

We consider a bipartite CV system comprising modes $j=1,2$, each described by
its respective bosonic annihilation and creation operator $\hat{a}_j$ and $\hat{a}_j^\dag$, such that $[\hat{a}_j,\hat{a}^\dag_{k}]=\delta_{jk}$ with $\delta_{jk}$ the Kronecker symbol being $1$ for $j=k$ and $0$ otherwise. For the aims of this work, it is worth introducing the quadrature operators $\hat{x}_j=\hat{a}_j+\hat{a}_j^\dag$ and $\hat{y}_j=i(\hat{a}^\dag_j-\hat{a}_j)$, which are collected in the vector of field quadratures $\hat{\bf Q}=(\hat{x}_1~\hat{y}_1~\hat{x}_2~\hat{y}_2)$.

A generic two-mode state with density matrix $\rho_{12}$ can be completely described in phase space by its Weyl characteristic function \cite{barnettradmore}
\begin{equation}
\chi (\xi,\eta) = \,{\text{Tr}}\,[\rho_{12} \hat D_1(\xi) \hat D_2(\eta)] \,, \label{cfs}
\end{equation}
 where  $\xi=\xi_r+i\xi_i$ and $\eta=\eta_r+i\eta_{i}$ are the phase-space variables, and we have introduced the phase-space displacement operator $\hat{D}_j(\alpha)=\exp[\alpha\hat{a}_j^\dag-\alpha^*\hat{a}_j]$ ($\alpha\in\mathbb{C}$).
It is straightforward to show that $\rho_{12}$ is explicitly related to the characteristic function as~\cite{glauber}
\begin{equation}
\label{density}
\rho_{12}=\frac{1}{\pi^2}\int{d}^2\xi{d}^2\eta~\chi(\xi,\eta)\hat{D}_{1}(-\xi)\otimes\hat{D}_{2}(-\eta).
\end{equation}
The exact characterization of a generic CV state requires in principle the knowledge of all the infinitely-many moments of the quadrature operators. The first moments can be adjusted by local displacements and are thus totally irrelevant for the purposes of characterizing and processing entanglement (they are assumed to be set to zero throughout this paper without loss of generality). On the other hand, second quadrature moments are often crucial: they can be conveniently collected in the covariance matrix $\bm V$ which, for a two-mode system, 
has entries $v_{ij}=\text{Tr}[\rho_{12}\{\hat{Q}_{i},\hat{Q}_{j}\}/2]$.
The covariance matrix of any physical two-mode state can be cast, via local unitary operations (which do not affect entanglement by definition), into the {\it standard form}~\cite{duan}
\begin{equation}
\label{standardform}
{\bm V}=
\begin{pmatrix}
{\bm n}_1&{\bm m}\\
{\bm m}&{\bm n}_2
\end{pmatrix}
\end{equation}
where each of the two block matrices ${\bm n}_j=\text{Diag}[n_j,n_j]$ accounts for the statistical properties of the corresponding mode $j$ ($n_{j}\ge{1}$), and ${\bm m}=\text{Diag}[m_+,m_-]$ specifies the cross-mode correlations ($m_+\ge|{m}_-|\ge{0}$). From now on, unless otherwise specified, we assume that the covariance matrix of any state at hand has already been put in standard form.

A very special role in the CV arena is played by Gaussian states. Gaussian states are defined as having a Gaussian characteristic function \cite{adessoilluminati}
\begin{equation}
\label{defi}
\chi(\xi,\eta)=e^{-\frac{1}{2}{\bf Q}\,{\bm V}\,{\bf Q}^T}\,,
\end{equation}
with ${\bf Q}=(\xi_r\,\xi_i,\eta_r\,\eta_i)$ the vector of canonical
phase-space variables.
Therefore, an arbitrary two-mode Gaussian state $\rho_{12}$ is completely specified (up to the first moments) by assigning the covariance matrix $\bm V$. The density matrix is directly expressed as a function of the covariance matrix via Eqs.~(\ref{density}) and (\ref{defi}). This greatly simplifies the mathematical treatment of Gaussian states, which despite living in a infinite-dimensional Hilbert space are effectively described by a finite number of degrees of freedom.

Physically, bipartite entanglement in a generic CV state is ``signalled'' by the violation of the Heisenberg uncertainty principle by the partially transposed density matrix~\cite{NPT,adessoilluminati}. For arbitrary non-Gaussian states, such an inseparability condition can be cast in terms of a hierarchy of inequalities involving moments of arbitrary order \cite{schuck}. In the special case of Gaussian states, clearly, entanglement is completely qualified and quantified by algebraic combinations of the elements of the covariance matrix ${\bm V}$ only. Specifically, a two-mode Gaussian state is entangled if and only if the following inequality is fulfilled,
\begin{equation}
\label{num}
\nu_- \equiv \frac{1}{\sqrt 2}\sqrt{\Delta-\sqrt{\Delta^2-4\det{\bm V}}}\le{1}
\end{equation}
with $\Delta=\det{\bm n}_1+\det{\bm n}_2-2\det{\bm m}$~\cite{adessoilluminati,serafini}.
In general, Eq.~(\ref{num}) stands as a sufficient condition for entanglement in an arbitrary non-Gaussian state with covariance matrix $\bm V$.

\begin{figure}[b]
\psfig{figure=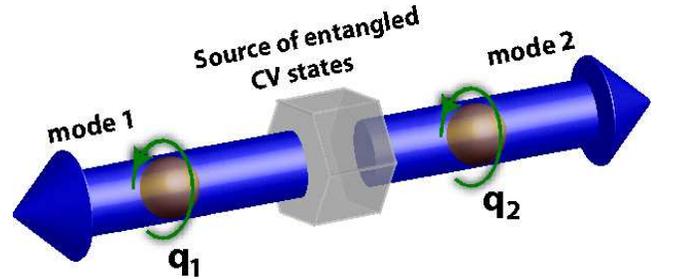,width=8.5cm,height=3.6cm}
\caption{(Color online) Scheme of principle for an entanglement transfer process. An entangled two-mode CV state is generated with an off-line process. Each mode addresses a two-level system $q_j~(j=1,2)$. Through bilocal interactions (represented by the circular arrows in the figure), part of the entanglement initially shared by modes $1$ and $2$ is transferred to the joint state of systems $q_1$ and $q_2$. The CV modes can then be discarded or can enter a detection stage for outcome post-selection, like in the {entanglement reciprocation} scheme~\cite{reciproco}.}
\label{figurina}
\end{figure}

\section{Bipartite Gaussian-state case}
\label{gaussiano}

In this Section we will derive a complete formal description of the entanglement transfer process from two-mode Gaussian states to two-qubit systems in terms of the covariance matrix of the Gaussian resource.
Eq.~(\ref{density}) is the starting point of the present analysis: we now use twice the decomposition of the identity in terms of Fock states $\sum^\infty_{n,m=0}\proj{n,m}{12}{n,m}=\openone$ and utilize it in order to get the Fock-state decomposition $\rho_{12}={\sum^{\infty}_{n,m=0}\sum^{\infty}_{p,q=0}}\gamma^{pq}_{nm}\proj{n,m}{12}{p,q}$ with the coefficients
\begin{equation}
\gamma^{pq}_{nm}=\frac{1}{\pi^2}\int{d}^2\xi{d}^2\eta~\chi(\xi,\eta)f_{np}(\xi)f_{mq}(\eta),
\end{equation}
where $f_{np}(\xi)=_1\!\langle{n}|\hat{D}_1(-\xi)|p\rangle_1$. Following Refs.~\cite{barnettradmore,glauber,oliveira}, one has that for $n\ge{p}$
\begin{equation}
f_{np}(\xi)=\sqrt{\frac{p!}{n!}}(-\xi)^{n-p}e^{-|\xi|^2/2}{\cal L}^{(n-p)}_p(|\xi|^2)
\end{equation}
with ${\cal L}^{(r)}_s(x)$ an associated Laguerre polynomial of degree $s$ and argument $x$ ($r\in\mathbb{Z}$). Analogous explicit forms hold for $f_{mq}(\eta)$. Although the four-fold summation involved in the full decomposition of $\rho_{12}$ seems daunting, we now show that, once the interaction model between the remote qubits and the two-mode CV system is assigned, the entanglement transfer scheme allows for some simplifications that considerably reduce the computational difficulties.

For the sake of definiteness, we assume that the two-level systems are both prepared in their fundamental energy level, associated with the logical state $\ket{g}_{q_j}$ ($j=1,2$). Any other choice is, obviously, equally valid. Then, in order to remain along the lines of the research conducted so far on entanglement transfer, we consider bilocal Hamiltonian models of the form
\begin{equation}
\label{model}
\hat{H}_{j,q_j}=\frac{\Omega}{2}(\hat{x}_j\hat{\sigma}^x_{q_j}+\hat{y}_j\hat{\sigma}^y_{q_j})~~~~~(j=1,2)
\end{equation}
with the $x,y$-Pauli matrix $\sigma^{x,y}_{q_j}$ and the coupling rate $\Omega$, which we take to be the same for both qubit-mode subsystems (again, an assumption that can be easily relaxed). Eq.~(\ref{model}) embodies a resonant dipole-like coupling Hamiltonian under the standard rotating wave approximation, an interaction mechanism of broad relevance that has attracted a considerable body of theoretical and experimental work in the last fifty years~\cite{JC}. We stress that the validity of the arguments that will be presented throughout this paper are not bound to this specific choice. Any coupling Hamiltonian could well be taken, {\it mutatis mutandis}, without affecting the main implications of the present study. The full form of the time-propagator $e^{-i\hat{H}_{j,q_j}t}$, decomposed in the single-qubit basis $\{\ket{e}_{q_j},\ket{g}_{q_j}\}$, has been given in~\cite{phoenix}. Here, it is sufficient to state that when exactly $n$ photons populate mode $j$, one has $\ket{g,n}_{q_j,j}\rightarrow{C}_{n}(\tau)\ket{g,n}_{q_j,j}-i{S}_{n}(\tau)\ket{e,n-1}_{q_j,j}$ with $\tau=\Omega{t}$, $C_{n}(\tau)=\cos({\tau}\sqrt{n})$ and $S_{n}(\tau)=\sqrt{1-C^2_n(\tau)}$. Equipped with these tools, one can now track the effective evolution of the two qubits obtained upon partial trace over the CV degrees of freedom. More formally
\begin{equation}
\rho_{q_1q_2}(\tau)=\text{Tr}_{12}[e^{-i\sum_{j}\hat{H}_{j,q_j}\tau}\rho_{12}\otimes\proj{gg}{q_1q_2}{gg}e^{i\sum_{j}\hat{H}_{j,q_j}\tau}].
\end{equation}

A simple calculation allows us to determine the $\gamma^{pq}_{nm}$ coefficients that are associated with each entry of the two-qubit density matrix $\rho_{q_1q_2}(\tau)$. One finds the relevant correspondences presented in Table~\ref{table:correspondence}, whose form can be verified upon explicit calculation (the Hermitian conjugates entries are easily found). The entries of Table~\ref{table:correspondence} are the only coefficients that are not identically null. This leaves us with the following two-qubit density matrix
\begin{equation}
\label{reduced}
\rho_{q_1q_2}(\tau)\!=\!\sum^\infty_{n,m=0}
\begin{pmatrix}
{\cal A}_{nm}(\tau)&0&0&{\cal G}_{nm}(\tau)\\
0&{\cal B}_{nm}(\tau)&{\cal D}_{nm}(\tau)&0\\
0&{\cal D}_{nm}(\tau)&{\cal C}_{nm}(\tau)&0\\
{\cal G}_{nm}(\tau)&0&0&{\cal E}_{nm}(\tau)
\end{pmatrix}
\end{equation}
with ${\cal E}_{nm}(\tau)=1-{\cal A}_{nm}(\tau)-{\cal B}_{nm}(\tau)-{\cal C}_{nm}(\tau)$. The form of $\rho_{12}(\tau)$ is significant. The coherence term $\sum_{n,m}{\cal G}_{nm}(\tau)$ arises in virtue of symmetric processes where both the qubits simultaneously absorb or emit an excitation, being thus responsible of the transformation $\ket{gg}_{q_1q_2}\leftrightarrow\ket{ee}_{q_1q_2}$. On the other hand, ${\cal D}_{nm}(\tau)$ accounts for anti-symmetric events where while $q_1$ emits or absorbs an excitation, $q_2$ undergoes the opposite physical process, thus giving rise to the $\ket{ge}_{q_1q_2}\leftrightarrow\ket{eg}_{q_1q_2}$ transition. Such parity-related effects are reflected into the apices and pedices of the associated $\gamma^{pq}_{nm}$ coefficients, whose form is still to be determined. This is the task of the following analysis.

\begin{table}[b]
\caption{Coefficients entering $\rho_{q_1q_2}(\tau)$. Not reported are the coefficients that turn out to be identically null upon explicit evaluation. For simplicity of notation, we have dropped any $\tau$-dependence.}
\centering
\begin{tabular}{c c}
\hline
\hline
Density-matrix entry & Coefficient \\
\hline
\hline
$\proj{gg}{q_1q_2}{gg}$ & ${\cal A}_{nm}\!=\!{C^2_nC^2_m}\gamma^{nm}_{nm}$\\
$\proj{ge}{q_1q_2}{ge}$ & ${\cal B}_{nm}\!=\!{C^2_nS^2_{m+1}}\gamma^{nm+1}_{nm+1}$\\
$\proj{eg}{q_1q_2}{eg}$ & ${\cal C}_{nm}\!=\!{S^2_{n+1}C^2_{m}}\gamma^{n+1m}_{n+1m}$\\
$\proj{gg}{q_1q_2}{ee}$ & ${\cal G}_{nm}\!=\!-C_nS_{n+1}C_{m}S_{m+1}\gamma^{n+1m+1}_{nm}$\\
$\proj{ge}{q_1q_2}{eg}$ & ${\cal D}_{nm}\!=\!{C}_nS_{n+1}C_{m}S_{m+1}\gamma^{n+1m}_{nm+1}$\\
\hline
\hline
\end{tabular}
\label{table:correspondence}
\end{table}

\begin{figure}[b]
{\bf (a)}\hskip4cm{\bf (b)}
\centerline{\psfig{figure=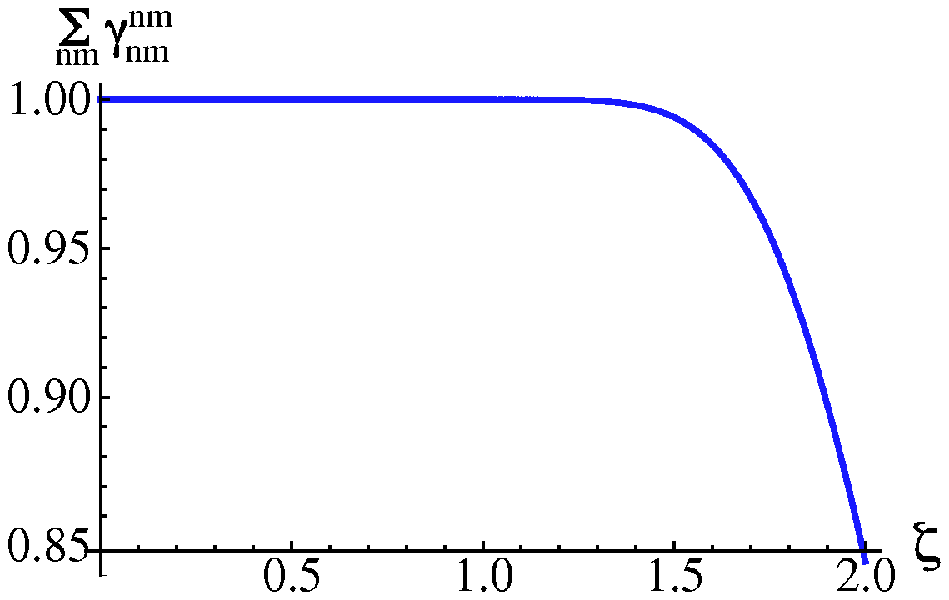,width=3.5cm,height=2.4cm}\psfig{figure=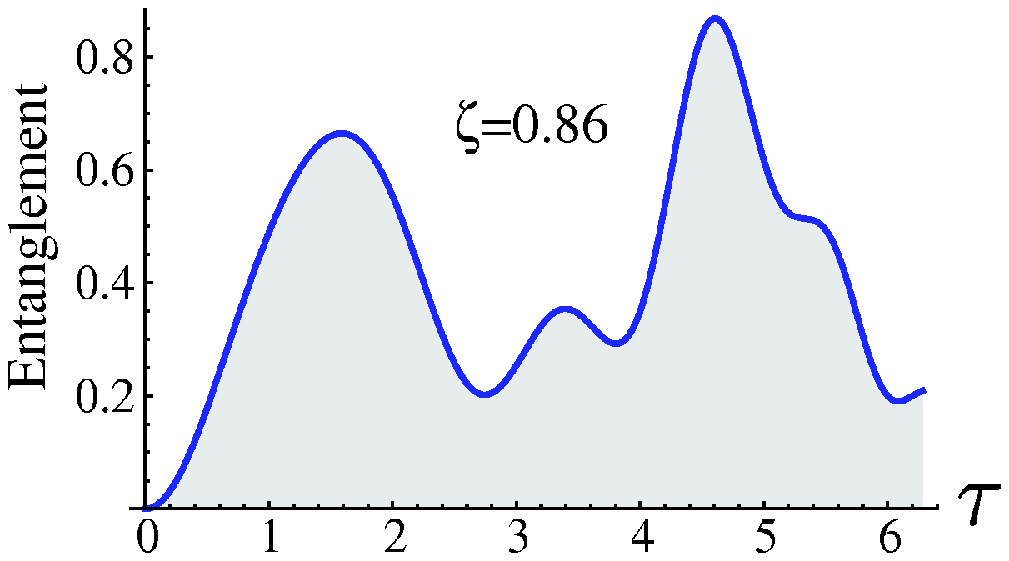,width=4.6cm,height=2.9cm}}
\caption{(Color online) {\bf (a)}: Normalization of a two-mode squeezed state evaluated as $\sum^{n_c}_{n}\sum^{m_c}_{m}\gamma^{nm}_{nm}$, shown against the squeezing parameter $\zeta$. The calculations are performed at fixed cutoffs $k_{c}=100$ and $n_{c}=m_c=25$. {\bf (b)}: Entanglement passed to two remote qubits upon bilocal interaction with a pure two-mode squeezed state of squeezing factor $\zeta=0.86$, plotted against the dimensionless time $\tau=\Omega{t}\in[0,2\pi]$. The calculation has been performed retaining the cutoffs given in panel {\bf (a)}. The result perfectly matches the studies in Refs.~\cite{ET}.}
\label{cutoff&test}
\end{figure}

It is important to stress that $\gamma^{pq}_{nm}$'s are numbers depending merely on the statistical properties of the two-mode CV systems via the Weyl characteristic function $\chi(\xi,\eta)$. As such, they can be written as explicit functions of the elements of the covariance matrix ${\bm V}$ associated with a given Gaussian CV state. Here, we highlight the relevant technical steps required for such calculations. In order to discuss a significant example, we consider the case of $\gamma^{nm}_{nm}$. First, one can conveniently change variables as $\xi=re^{i\phi},\eta=se^{i\theta}$ (with $r,s\in[0,\infty[$ and $\phi,\theta\in[0,2\pi]$). The four-fold integral entering the definition of $\gamma^{pq}_{nm}$ is thus changed into a double integral
\begin{equation}
\begin{aligned}
\gamma^{nm}_{nm}&=\frac{1}{\pi^2}\int^\infty_0r\,dr\,\int^\infty_0s\,ds\,e^{-\frac{1}{2}(r^2+s^2)}{\cal L}^{(0)}_{n}(r^2){\cal L}^{(0)}_{m}(s^2)\\
&\times\int^{2\pi}_{0}d\phi\int^{2\pi}_{0}d\theta~\chi(re^{i\phi},se^{i\theta}).
\end{aligned}
\end{equation}
The double integral involving the angular variables can be worked out by writing explicitly the form of the characteristic function and using a power-series expansion. We end up with $\int^{2\pi}_0{d}\phi\!\int^{2\pi}_{0}{d}\theta~\chi(re^{i\phi},se^{i\theta})=4\pi^2e^{-\frac{n_1r^2+n_2s^2}{2}}\sum^\infty_{k=0}(rs)^{2k+1}G_{k}(m_\pm)$ with
\begin{equation}
G_{k}(m_\pm)=\frac{(2k-1)!!}{(2k)!(2k)!!}m^{2k}_{-}\,{}_2F_1\left(\frac{1}{2},-k;1;\frac{m^2_--m^2_+}{m^2_-}\right),
\end{equation}
where ${}_2F_1$ is the Gauss hypergeometric function. The remaining {\it radial} part of the integral is easily evaluated by exploiting the power-series decomposition of the associated Laguerre polynomials~\cite{AS} and the Gaussian integral $\int^\infty_0dr{e}^{-\frac{r^2}{2}}r^{2k+2i+1}=(k+i)!/2~\forall{i}\in\mathbb{Z}$~\cite{GR}. After some simplifications, one gets the final expression
\begin{equation}
\label{pergggg}
\begin{aligned}
&\gamma^{nm}_{nm}\!=\!\sum^\infty_{k=0}\frac{4m^{2k}_-}{[(n_1+1)(n_2+1)]^{k+1}}\,{}_2F_1\left(-n,1+k;1;\frac{2}{n_1+1}\right)\\
&\times{}_2F_1\left(-m,1+k;1;\frac{2}{n_2+1}\right){}_2F_1\left(\frac{1}{2},-k;1;\frac{m^2_--m^2_+}{m^2_-}\right).
\end{aligned}
\end{equation}
Any other $\gamma^{pq}_{nm}$ involved in $\rho_{q_1q_2}(\tau)$ can be calculated following similar steps, which we omit here. In Appendix~A we give the explicit form of all the relevant ones. It is worth stressing that, regardless of the apparently intimidating form of the coefficients in Eqs.~(\ref{pergggg}) and Eqs.~(\ref{pergege})-(\ref{pergeeg}), they are rather useful and pragmatic. Their {\it plug-\&-play} nature allows an unprecedented flexibility in the study of entanglement transfer. It is enough to assign the covariance matrix of the two-mode Gaussian resource, without the necessity of knowing the full decomposition of the state onto a given basis, in order to  unambiguously determine the amount of entanglement that can be transferred to two remote qubits prepared in their ground state. In the next Sections we demonstrate the flexible nature of our results by addressing significant quantitative examples. It is usually the case that the infinite sums involved in the determination of $\rho_{q_1q_2}(\tau)$ can be truncated using appropriate cutoffs, their value depending on the features of ${\bm V}$. For instance, for a two-mode squeezed state of squeezing factor $\zeta\in{\mathbb R}$ we have $n_{1,2}=\cosh(2\zeta),m_\pm=\pm\sinh(2\zeta)$. Obviously, $\text{Tr}_{12}(\rho_{12})=1$, which translates into $\sum_{n,m}\gamma^{nm}_{nm}=1$, as it is easy to check. In order to explicitly evaluate the normalization of the state, we have truncated the sum over $k$ in Eq.~(\ref{pergggg}) and analogous at $k_c=100$ while those over $n$ and $m$ at $n_c=m_c=25$. These cutoffs provide excellent results for values of $\zeta$ up to $\sim1.5$, as seen in Fig.~\ref{cutoff&test} {\bf (a)}. Larger squeezing parameters require larger $n_c$ and $m_c$ values, as it is intuitive to understand. It is well-known, in fact, that the probability that the Fock state with $n$ photons is populated in a squeezed state of $\zeta\gtrsim{1}$ is non-negligible for quite a large range of $n$ values.

The validity of the results presented so far can be conveniently tested by using this very same instance of two-mode CV resource for entanglement transfer and comparing the outcomes with available literature on the subject~\cite{ET}. Throughout this work, we use the entanglement measure based on the negativity of partial transposition criterion~\cite{NPT}, which is defined as twice the modulus of the negative eigenvalue of the partially-transposed two-qubit density matrix with respect to one of the two qubits~\cite{measure}. Using the coefficients calculated here and setting $\zeta=0.86$, which is known to optimize the entanglement transfer with this type of resource~\cite{ET}, we find the behavior shown in Fig.~\ref{cutoff&test} {\bf (b)}, which is in perfect quantitative agreement with the results of Refs.~\cite{ET} and thus states the reliability of our findings. It should be stressed that, as discussed in Ref.~\cite{ET}, the aperiodic behavior shown in Fig.~(\ref{cutoff&test}) {\bf (b)} results from the interference of pure Rabi oscillations, each occurring at frequency $\sqrt{n}\Omega$, associated with transitions within each two-level system induced by the chosen multi-photon resource. The incommensurability of the periods of such oscillations is responsible for the destruction of any periodicity.

\subsection{Thermal nature and dissipation effects in the CV resource}
\label{termale}

We now start discussing the remarkable flexibility of the established connection between transferred entanglement and statistical properties of the CV resource. Our first step is the use of a more general Gaussian resource than a standard two-mode squeezed state. The most general state of a single-mode CV system is given by a displaced squeezed thermal state. However, when the multi-mode scenario is considered, many possibilities of combining and ordering single-mode as well as two-mode unitary operations are available. In order to include a standard two-mode squeezed state as a limiting case, here we concentrate on the following situation: we consider single-mode squeezing (along arbitrary directions in phase space) of two CV modes in thermal equilibrium at their respective temperature. Such squeezed thermal modes are then superimposed at a beam splitter of transmittivity $T$. More formally, we consider the state
\begin{equation}
\label{thermalsqueezed}
\rho_{12}=\hat{B}(\hat{R}_1\otimes\hat{R}_2)(\hat{S}_{1}\otimes\hat{S}_{2})\rho^{th}_{12}(\hat{S}_{1}\otimes\hat{S}_{2})^\dag(\hat{R}_1\otimes\hat{R}_2)^\dag\hat{B}^\dag
\end{equation}
\begin{figure}[b]
\psfig{figure=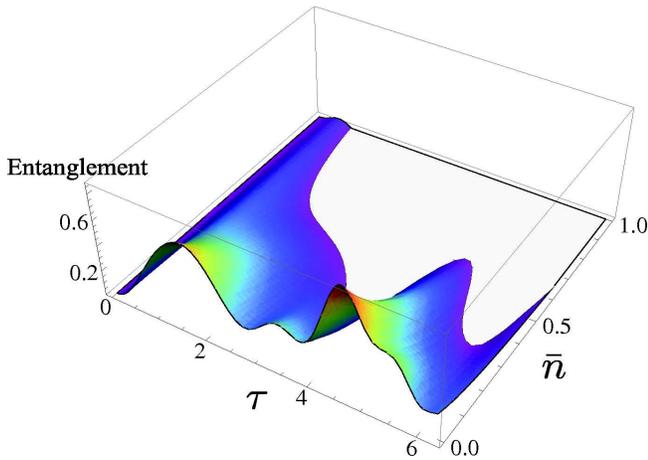,width=8.5cm,height=6cm}
\caption{(Color online) Transferred entanglement against dimensionless interaction time $\tau$ and the mean thermal occupation number $\nbar_{1,2}=\nbar$ of the two CV modes used as resources. For the whole range of $\nbar$ shown in this plot, the initial CV resouce is nonseparable.}
\label{termico}
\end{figure}
with $\hat{B}=\exp[i\cos^{-1}(\sqrt T)(\hat{x}_1\hat{y}_2-\hat{y}_1\hat{x}_2)]$ the beam splitter operator, $\hat{S}_{j}$ and $\hat{R}_j$ the phase-space single-mode squeezing and rotation operators~\cite{barnettradmore} (whose form is here omitted). We have introduced the tensor product of two thermal states $\rho^{th}_{12}=\sum^\infty_{n,m=0}\beta_{nm}\proj{n,m}{12}{n,m}$, which we take to have mean thermal occupation number $\nbar_{1}$ and $\nbar_2$ respectively, so that $\beta_{nm}=(\nbar^n_1\nbar^m_2)/(\nbar_1+1)^{n+1}(\nbar_2+1)^{m+1}$~\cite{barnettradmore}. Depending on the beam-splitter transmittivity and the relative direction of squeezing, the mixed state $\rho_{12}$ can show some entanglement. While its Fock-state decomposition is somehow uncomfortable and definitely lengthy to compute, it is immediate to find the form of the covariance matrix ${\bm V}$. Each of the unitaries in Eq.~(\ref{thermalsqueezed}) has, indeed, a symplectic counterpart that can be used to transform the covariance matrix of the initial thermal state, which reads ${\bm V}_{\rho^{th}}=(2\nbar_1+1)\openone_2\oplus(2\nbar_2+1)\openone_2$ with $\openone_k$ the ${k}\times{k}$ identity matrix. One has $S_j=\text{Diag}[e^{-s_j},e^{s_j}]$, $R_{j}=\cos\varphi_j\openone_2+i\sin\varphi_j\sigma_y$ and $B=\begin{pmatrix}\sqrt{T}\openone_2&-\sqrt{1-T}\openone_2\\\sqrt{1-T}\openone_2&\sqrt{T}\openone_2\end{pmatrix}$. Here, $s_j$ and $\varphi_j$ are the single-mode squeezing parameter and phase-space rotation angle, respectively. The covariance matrix thus changes as
\begin{equation}
\label{cv}
{\bm V}_{\rho_{12}}=B^T(R_1\oplus{R}_2)^T(S_1\oplus{S}_2)^T{\bm V}_{\rho^{th}}(S_1\oplus{S}_2)(R_1\oplus{R}_2)B.
\end{equation}
Eq.~(\ref{cv}), which is readily calculated upon simple matrix products, is then put in standard form by following the general recipe given in Ref.~\cite{duan}.
With this, the evaluation of the transferred entanglement for a given choice of the parameters entering Eq.~(\ref{cv}) is immediate. In order to study the effects that mixedness in the CV resource has in the transfer performance, a step which has only been superficially addressed so far, here we discuss the case of $s_1=-s_2=0.86$, $\varphi_{1,2}=0$, $T=1/2$ and $\nbar_{1,2}=\nbar$. Interesting effects arising from the use of asymmetric two-mode Gaussian resources ({\it i.e.} $\nbar_1\neq{\nbar_2}$) will be addressed elsewhere~\cite{noi}. With these choices, for $\nbar=0$ we have the covariance matrix of a two-mode squeezed state and we thus retrieve the results of Fig.~\ref{cutoff&test} {\bf (b)}. At non-zero thermal occupation number, on the other hand, we have $n_{1,2}=(1+2\nbar)\cosh(1.72)$ and $m_\pm=\pm(1+2\nbar)\sinh(1.72)$, which lead to Fig.~\ref{termico}. The amount of transferred entanglement decreases with $\nbar$, as it should be expected given that the entanglement within the CV resource itself is spoiled by an increasing mixedness $\nbar\neq{0}$. However, the relevant feature is that we find large temporal regions of separability of the qubits' state, despite the CV resource remains entangled within the full range of values of $\nbar$ shown in the plot. In fact, we get $\nu_-=1$ (cfr. Eq.~(\ref{num})) at $\nbar=(e^{2\zeta}-1)/2$, which is equal to $2.292$ for  $\zeta=0.86$. The thermal nature of the CV state thus makes the entanglement-transfer process unreliable.

\begin{figure}[t]
\centerline{\psfig{figure=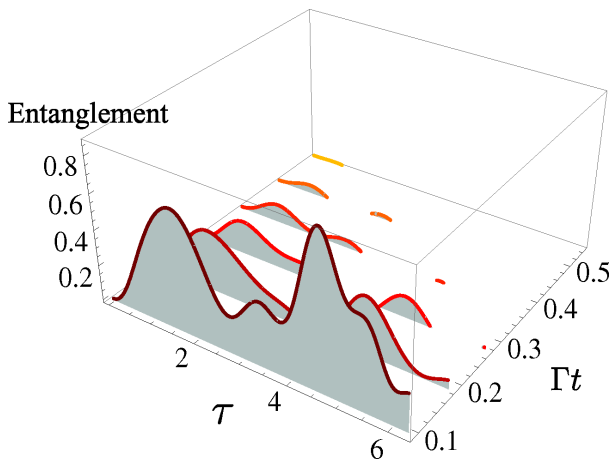,width=8.0cm,height=6.0cm}}
\caption{(Color online) Qubit entanglement transferred from a two-mode squeezed state subject to dissipation, plotted against the dimensionless interaction time $\tau$ and dissipation time $\Gamma{t}$ for $N=0.1$ and $\zeta=0.86$. }
\label{dissipazione}
\end{figure}

On the other hand, one can also consider the effect that dissipation affecting the CV resource has on the passage of quantum correlations. Again, having related the transferred entanglement to the elements of the covariance matrix of the two-mode state proves to be a major advantage. In fact, by assuming weak coupling between each mode $j=1,2$ and its respective bosonic bath at thermal equilibrium, which allows our analysis to be kept within the Born-Markov approximation, the dissipative dynamics of the CV system is described by the master equation~\cite{wallsmilburn}
\begin{equation}
\begin{aligned}
\partial_{t}\rho_{12}&=\frac{\Gamma}{2}\sum^2_{j=1}\left[N(2\hat{a}^\dag_j\rho_{12}\hat{a}_j-\{\hat{a}_j\hat{a}^\dag_j,\rho_{12}\})\right.\\
&\left.+(N+1)(2\hat{a}_j\rho_{12}\hat{a}^\dag_j-\{\hat{a}^\dag_j\hat{a}_j,\rho_{12}\})\right]
\end{aligned}
\end{equation}
with $\Gamma^{-1}$ the lifetime of a photon in the dissipative environment characterized by its mean thermal number $N$. By making proper use of standard operator correspondences~\cite{wallsmilburn}, this master equation can be changed into a dynamical equation for the characteristic function $\chi(\xi,\eta)$. After a tedious but otherwise straightforward calculation, one gets
\begin{equation}
\label{chareq}
\partial_t\chi(\xi,\eta)\!=\!-\frac{\Gamma}{2}\sum^2_{j=1}[(1+2N)|\mu_j|^2+\mu^*_j\partial_{\mu_j}+\mu_j\partial_{\mu^*_j}]\chi(\xi,\eta)
\end{equation}
with $\mu_1=\xi$ and $\mu_2=\eta$. Eq.~(\ref{chareq}) is readily solved and, by means of definition (\ref{defi}), it is finally found that the initial covariance matrix ${\bm V}(0)$ evolves towards the covariance matrix of the bath $(2N+1)\openone_4$ as
\begin{equation}
\label{decoCV}
{\bm V}(t)=(2N+1)(1-e^{-\Gamma{t}})\openone_4+{\bm V}(0)e^{-\Gamma{t}},
\end{equation}
which is in agreement with the analysis reported in Refs.~\cite{serale}. We now plug the covariance matrix of a two-mode squeezed state as ${\bm V}(0)$ in Eq.~(\ref{decoCV}) and study the modifications induced in the entanglement transfer function, for set values of $N$ and $\zeta$, by an increasing ``dissipation time" $\Gamma{t}$. This dimensionless parameter is the product of the dissipation rate $\Gamma$, which characterizes the environmental action, and the time $t$ dissipation has acted on the CV resource before the entanglement transfer process begins~\cite{nota}. The results are shown in Fig.~\ref{dissipazione}, where $\zeta=0.86$ is taken with $N=0.1$. Quite expectedly, the transferred entanglement decreases with $\Gamma{t}$ and disappears for $\Gamma{t}\simeq{0.5}$. This matches what is found for the two-mode CV resource, whose $\nu_-$ becomes larger than $1$ for
\begin{equation}
\Gamma{t}>\log \left(\sqrt{\frac{4{N}(N+1)+\sinh (2 r)-\cosh
   (2 r)+1}{4N^2+4N}}\right),
\end{equation}
which equals $0.52$ for the choice of parameters made above. One can thus claim that the entanglement transfer process is effective as far as the CV resource is entangled, although regions of separability appear, dynamically, within the temporal window studied in Fig.~\ref{dissipazione}. This behavior has been checked to hold true also when the initial two-mode covariance matrix describes a general physically-allowed state other than a two-mode squeezed vacuum.

\section{Bipartite non-Gaussian resources}
\label{nongaussiano}

The handy nature of the connection established between statistical properties of the entangled resource and transferred two-qubit entanglement turns out to be even more striking when non-Gaussian CV states are at hand. In this Section, we show that the initial Gaussian component of a de-Gaussified two-mode state is key in the determination of the entanglement shared by the remote two-level systems.
We consider de-Gaussification as obtained by photon-subtraction from a general Gaussian CV state~\cite{refs,kitagawa}.

An intuition of our claim comes from considering the definition of an $s$-photon subtracted two-mode CV state, which reads
\begin{equation}
\label{sottrattoformale}
\rho'_{12}=\frac{{\cal N}}{\pi^2}\int{d}^2\xi{d}^2\eta~\chi(\xi,\eta)(\hat{a}_1^s\hat{D}_1(-\xi)\hat{a}_1^{\dag{s}})(\hat{a}_2^s\hat{D}_2(-\eta)\hat{a}_2^{\dag{s}})
\end{equation}
with ${\cal N}$ the normalization factor and $s$ the number of photons subtracted (symmetrically) from the two modes of the CV resource. Although this equation is only formal, it encompasses the crucial features entailed by the physical de-Gaussification process of subtracting a photon. In Appendix~B
we also highlight the results achieved upon use of an effective procedure for subtracting photons, making use of highly-biased beam-splitters and  photo-resolving detectors.

It is straightforward to check that the determination of the coefficients associated with the various elements of the two-qubit density matrix resulting from the interaction with the non-Gaussian CV resource tracks the steps depicted in Sec.~\ref{gaussiano} and Appendix~A. In fact, all one has to do is to take
\begin{equation}
\label{modificati}
\gamma^{pq}_{nm}\rightarrow\sqrt{\frac{(n+s)!(m+s)!(p+s)!(q+s)!}{n!m!p!q!}}\gamma^{p+s\,q+s}_{n+s\,m+s}
\end{equation}
in the entries of Tab.~\ref{table:correspondence} and matrix $\rho_{q_1q_2}$. The presence of the square-root factor reflects the effects of de-Gaussification process. This result demonstrates our claim that the entanglement-transfer capabilities of an $s$-photon subtracted state, under the model for local Hamiltonian addressed within the context of this work, are determined by the knowledge of the Gaussian ``core" of the state and, of course, the number of photons being subtracted.

\begin{figure}[b]
\centerline{\psfig{figure=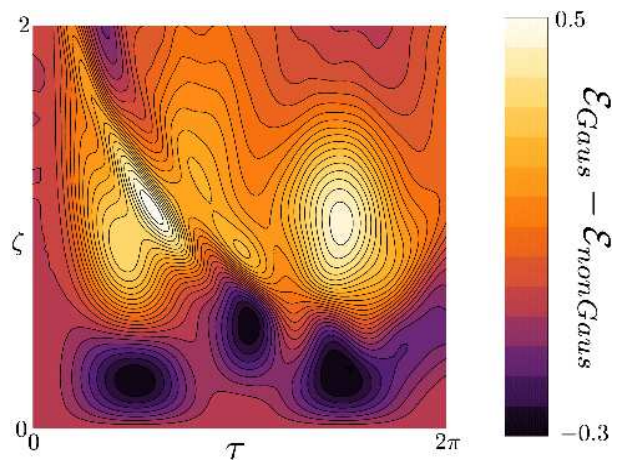,width=8.0cm,height=6.0cm}}
\caption{(Color online) Density plot of the difference ${\cal E}_{Gaus}-{\cal E}_{nonGaus}$ between the entanglement transferred using a two-mode squeezed state and that achieved via a $1$-photon subtracted two-mode squeezed state. The transferred entanglement is studied against the dimensionless interaction time $\tau$ and the squeezing parameter $\zeta$. The side color-bar indicates the color scale.}
\label{differenza}
\end{figure}

\begin{figure}[t]
\centerline{\psfig{figure=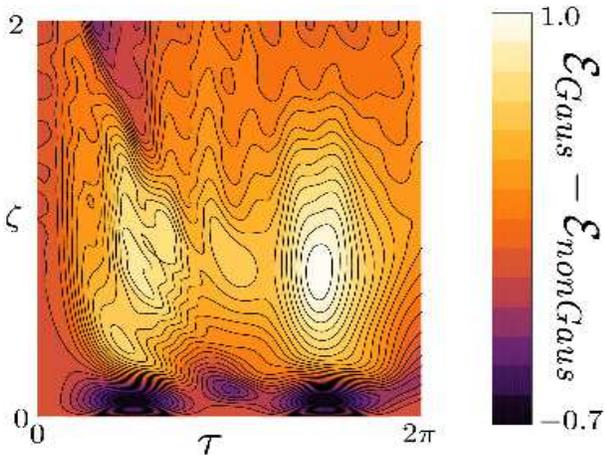,width=8.0cm,height=6.0cm}}
\caption{(Color online) Density plot of the difference ${\cal E}_{Gaus}-{\cal E}_{nonGaus}$ between the entanglement transferred using a two-mode squeezed state and that achieved via a $7$-photon subtracted two-mode squeezed state. The transferred entanglement is studied against the dimensionless interaction time $\tau$ and the squeezing parameter $\zeta$. The side color-bar indicates the color scale.}
\label{differenzadopo7}
\end{figure}

\begin{figure}[b]
\centerline{\psfig{figure=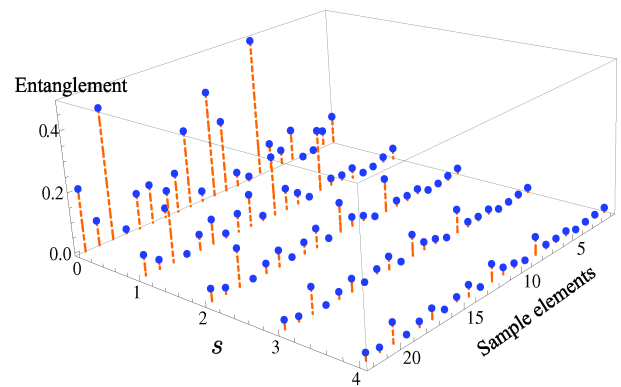,width=8.0cm,height=5cm}}
\caption{(Color online) We show the maximum amount of entanglement transferred within $\tau\in[0,2\pi]$ from a sample of 22 randomly generated $s$-photon subtracted Gaussian state for $s=0,..,4$. As more photons are subtracted from the Gaussian core part of the state, despite an increasing resource entanglement, the maximum quantum correlations that can be passed to two remote qubits decreases.}
\label{maxdecrease}
\end{figure}

Our task now becomes manyfold. First, we aim at showing that a non-Gaussian resource obtained from a two-mode squeezed state, which is a realistic and interesting case to study~\cite{kitagawa},  beats the corresponding Gaussian resource only for proper choices of the squeezing parameter $\zeta$. However, the maximum value of entanglement transferred by a de-Gaussified state for $\tau\in[0,2\pi]$, which is a reasonable time to wait, never exceeds the one achieved via a Gaussian resource. This strongly contrasts with the entanglement of the state itself: the negativity of an $s$-photon subtracted two-mode squeezed state is larger than the one for $s=0$, regardless of $\zeta\ge0$ and $s>0$. In the remainder of this Section, we indicate with ${\cal E}_{Gauss}$ (${\cal E}_{nonGaus}$) the transferred negativity when a Gaussian (de-Gaussified) CV resource is used.  
In Fig.~\ref{differenza} we show the difference between the entanglement ${\cal E}_{Guas}$ transferred upon usage of a Gaussian two-mode squeezed state and ${\cal E}_{nonGaus}$ when a $1$-photon subtracted two-mode squeezed state is employed. Such a difference is evaluated at set values of the dimensionless interaction time $\tau$ and the degree of squeezing of the Gaussian-core resource. We notice a rich structure of maxima and minima. Notably, with the exception of the three dark areas in the region of small $\zeta$, witnessing that the use of the non-Gaussian resource favors the transfer of entanglement to the remote two-level systems, such a difference remains largely non-negative throughout the entire region determined by $\tau\in[0,2\pi]$ and $\zeta\in[0,2]$. The results are thus clear: the use of a Gaussian CV resource is (almost always) much more effective than a non-Gaussian state (belonging to the class of de-Gaussified states studied here) when entanglement transfer processes are addressed. This qualitative behavior holds for any value of $s$ tested by our calculations. For instance, for $s=7$ we get the plot in Fig.~\ref{differenzadopo7}, which shows that the three negative areas where the use of a de-Gaussified state is more advantageous get squashed in the low-squeezing region, while the moduli of ${\cal E}_{Gaus}-{\cal E}_{nonGaus}$ increase, thus showing an even more striking convenience in using Gaussian states. The additional square-root factor in Eq.~(\ref{modificati}), which deeply affects the interferences among components associated with fixed numbers of excitations, is the reason for such a considerable change in behavior of the discrete-variable entanglement function. The squashing towards regions of lower squeezing originates from the modifications induced by photon-subtraction on the photon-number probability distributions of the states being used: as $s$ grows, such probability distributions are peaked at smaller $\zeta$ and their width shrinks with respect to what occurs for the $s=0$ case, thus explaining the behavior observed in Figs.~\ref{differenza} and~\ref{differenzadopo7}. In order to show that  Gaussian states allow for the achievement of the largest transferred  entanglement, regardless of the form of the entangled resource, we have used the general formalism of  Sec.~\ref{termale} to generate a 1000-element sample of {\it bona fide} random covariance matrices. They have been used in order to evaluate the maximum qubit entanglement (for $\tau\in[0,2\pi]$) that can be achieved by using such fully Gaussian resources and the $s$-photon subtracted states having the latter as Gaussian core (with $s=1,..,4$). The result is that, quite clearly, Gaussian states are able to achieve the largest discrete-variable entanglement, as far as the model for bilocal interaction discussed here is considered. Moreover, the successive subtraction of photons reduces the maximum of entanglement being transferrable. These features are clearly shown, for a subset of only $22$ elements, in Fig.~\ref{maxdecrease}, where the transferred entanglement upon use of one of such elements is studied against the number of photon subtractions performed in the CV resource: the decrease with $s$, regardless of the Gaussian core part of the resource, is quite evident.

We now address a second interesting point related to the use of non-Gaussian CV states. General considerations in the theory of CV entanglement reveal that the Gaussian state having, as covariance matrix entries, the moments (up to the second) of a general two-mode CV state, provides a lower bound to the entanglement content of the latter. In this sense, Gaussian states are said to be ``extremal''~\cite{wolf}. Recently, it has been proven that the entanglement in an $s$-photon subtracted two-mode squeezed state is upper- and lower-bounded by a functional of the second moments of such non-Gaussian state~\cite{adesso}. Here, we show that this is actually not the case for the amount of transferred entanglement: the one achieved via such ``fictitious'' Gaussian resource (with equal first and second moments) is often larger than the entanglement passed to two remote qubits by the (in general non-Gaussian) CV state. This can be very easily checked by proceeding as follows. One can take the covariance matrix elements of an $s$-photon subtracted  two-mode squeezed state as given in Ref.~\cite{adesso}, use the expressions valid for the Gaussian-resource case (cfr. Eq.~(\ref{pergggg}) and the formulas in Appendix~A) and compare the results with what is gathered by considering the formal apparatus, presented earlier in this Section, for $s$ photons being subtracted from it. The explicit calculations confirm the general trend anticipated here that Gaussian CV resources appear to be optimal for entanglement transfer purposes. Moreover, it allows us to conclude that extremal states in the CV scenario are not mapped, in general, into extremal discrete-variable states by the entanglement transfer process at hand.

We are currently looking for a physically clear relationship involving such a discrepancy between a non-Gaussian state and its Gaussian-equivalent one and the {\it degree of non-Gaussianity} of the starting CV resource as quantified by quantum relative entropy~\cite{paris}. For the case of an $s$-photon subtracted two-mode squeezed state, such a figure of merit simply coincides with the von Neumann entropy of the Gaussian-equivalent state associated with it. We already have a few numerical evidences showing that the squeezing-dependent functional form of the difference between the maxima of transferred entanglement in the non-Gaussian and Gaussian-equivalent cases mimics the shape of the degree of non-Gaussianity in the low-squeezing region.

\section{Conclusions and perspectives}
\label{conclusioni}

We have provided a general formalism for the treatment of entanglement-transfer processes from an arbitrary two-mode resource to remote two-level systems. The remarkable flexibility of the method proposed here allows for the generalization and extension of this protocol for the distribution of long-haul quantum communication channels to situations that have been only partially addressed so far, such as the use of thermal/dissipated Gaussian CV resources and experimentally interesting de-Gaussified states~\cite{furusawa}.

The proposed methodology is quite versatile and can be straightforwardly adapted in a way so as to consider multi-mode CV resources, as in Ref.~\cite{altri1}, and include the effects of post-selection and detection, such as in the entanglement reciprocation scheme of Refs.~\cite{reciproco,reciproco2}. In this context, the entanglement transfer protocol~\cite{ET} as described and studied here is complemented by ``hard" projective measurements performed on the entangled resource. One can consider two complementary cases. In the first, a two-mode CV state is used to entangle two separate qubits by means of local interactions and post-selective measurements of a specific and appropriate nature. While Ref.~\cite{reciproco} considered the case of a projection of the CV resource onto coherent states of a proper amplitude~\cite{barnettradmore}, other choices, such as parity or homodyne measurements, are possible (see Ref.~\cite{tufarelli} for a recent example).  In general, upon projection of modes $1$ and $2$ by means of the operator $\hat{\Pi}_1\otimes\hat{\Pi}_2$
 such that $\hat{\Pi}^2_j=\hat{\Pi}_j~(j=1,2)$, it is immediate to see that the reduced density matrix of the two qubits is cast into the form
 \begin{equation}
 \label{reciproco}
 \rho_{q_1q_2}\!\propto\!\sum_{}\!\gamma^{pq}_{nm}\varrho^{np}_{q_1}(\hat{\Pi}_1,\tau)\varrho^{mq}_{q_2}(\hat{\Pi}_2,\tau)
\end{equation}
with $\varrho^{np}_{q_1}(\hat{\Pi}_1,\tau)$ [$\varrho^{mq}_{q_2}(\hat{\Pi}_2,\tau)$] a time-dependent operator spanning the Hilbert space of qubit $q_1$ ($q_2$) whose form, for an assigned interaction Hamiltonian, depends on the choice of $\hat{\Pi}_1$ ($\hat{\Pi}_2$) and on the initial preparation of the qubits. The coefficients $\gamma^{pq}_{nm}$ are calculated as described in the previous Sections. One can also consider the reverse situation where the state of qubits $q_1$ and $q_2$, entangled as a result of the process described above, is used as a resource to entangle two modes, labelled for simplicity $1$ and $2$, which are initially in a separable state. Upon proper projection of the qubits, an expression analogous to (\ref{reciproco}) would be obtained, with $\hat{\Pi}_{1,2}$ to be interpreted, this time, as qubit projectors. Clearly, the effectiveness of the reciprocation of entanglement would depend on the sort of projection being chosen. We conclude that the results discussed in this paper can be fully exploited in order to assess the case of entanglement reciprocation as well, which depend crucially on the very same coefficients $\gamma^{pq}_{nm}$ calculated in this work.

We expect that the handiness of the results achieved through our methodology will trigger further development and deepening of entanglement-transfer processes that, in light of recent experimental progresses along these very same lines~\cite{kimble}, hold the promises to embody a pragmatically viable route to long-haul distribution of quantum correlations.

\acknowledgments
We thank D. Ballester and C. Di Franco for discussions. MP dedicates this work to Sara Vittoria Paternostro. We acknowledge support from DEL and the UK EPSRC (EP/G004579/1).

\renewcommand{\theequation}{A-\arabic{equation}}
\setcounter{equation}{0}
\section*{APPENDIX A}
\label{espliciti}

In this Appendix we give the explicit form of the $\gamma^{pq}_{nm}$ coefficients involved in the
two-qubit density matrix resulting from the interaction with a CV system prepared in a Gaussian state.
In Eq.~(\ref{pergggg}) we have already provided the one associated with the $\proj{gg}{q_1q_2}{gg}$
projector. We start looking at the coefficient for $\proj{ge}{q_1q_2}{ge}$, which depends on $\gamma^{nm+1}_{nm+1}$. By definition
\begin{equation}
\label{pergege}
\begin{aligned}
&\gamma^{nm+1}_{nm+1}=\frac{1}{\pi^2}\int{d}^2\xi{d}^2\eta~\chi(\xi,\eta)f_{nn}(\xi)f_{m+1m+1}(\eta)\\
&=\sum^\infty_{k=0}\frac{4m^{2k}_-}{[(n_1+1)(n_2+1)]^{k+1}}{}_2F_1\left(-n,1+k;1;\frac{2}{n_1+1}\right)\\
&{}_2F_1\left(-m-1,1+k;1;\frac{2}{n_2+1}\right)\!{}_2F_1\left(\frac{1}{2},-k;1;\frac{m^2_--m^2_+}{m^2_-}\right).
\end{aligned}
\end{equation}
In a completely analogous way, we get that $\gamma^{n+1m}_{n+1m}$ is given by Eq.~(\ref{pergege}) with the replacement $n\rightarrow{n+1}$ and $m+1\rightarrow{m}$. As for the $\proj{gg}{q_1q_2}{ee}$
element we have
\begin{equation}
\label{perggee}
\begin{aligned}
&\gamma^{n+1m+1}_{nm}=
-\frac{4^{k+1} [(k+1)!]^2 \sqrt{(m+1) (n+1)}}{(2 k+1)![(n_1+1)(n_2+1)]^{k+2}}K_k(m_\pm)\\
&\times{}_2F_1\left(-n,2+k;2;\frac{2}{n_1+1}\right)
 {}_2F_1\left(-n,2+k;2;\frac{2}{n_2+1}\right)
\end{aligned}
\end{equation}
with $K_k(m_\pm)$ a combination of the hypergeometric functions ${}_2F_1\left(\pm\frac{1}{2},-k;2;1-\frac{m^2_+}{m^2_-}\right)$ with coefficients entirely determined by $m_\pm$.
A somehow analogous form holds for the coefficient associated with $\proj{ge}{q_1q_2}{eg}$, which reads
\begin{equation}
\label{pergeeg}
\begin{aligned}
&\gamma^{n+1m}_{nm+1}=-\frac{2^{2 k+3}[(k+1)!]^2 \sqrt{(m+1)(n+1)}}{(2 k+1)![(a+1) (b+1)]^{k+2} } L_k(m_\pm)\\
&\times{}_2F_1\left(-n,2+k;2;\frac{2}{a+1}\right){}_2F_1\left(-m,2+k;2;\frac{2}{b+1}\right).
\end{aligned}
\end{equation}
As before, $L_{k}(m_\pm)$ is fully determined by the correlation terms of the CV systems's covariance matrix and contains hypergeometric functions.

All the remaining density matrix elements are associated with coefficients that turn out to be identically null. In order to give an intuition for this, let us consider $\gamma^{nm+1}_{nm}$, which accounts for  $\proj{gg}{q_1q_2}{ge}$. The angular part of the four-fold integral to be performed reads
\begin{equation}
\int^{2\pi}_0d\phi\int^{2\pi}_0{d}\theta~\chi(re^{i\phi},se^{i\theta})e^{-i\theta}=0
\end{equation}
for any covariance matrix ${\bm V}$ in standard form as in Eq.~(\ref{standardform}). Similar arguments hold for any other identically-zero coefficient in $\rho_{q_1q_2}(\tau)$.

\renewcommand{\theequation}{B-\arabic{equation}}
\setcounter{equation}{0}
\section*{APPENDIX B}
\label{fisico}

In this Appendix we show the formal derivation of the coefficients valid for the case of a photon subtracted state, when the implementation of the non-Hermitian operator $\hat{a}_j$ occurs via highly-biased beam splitters and the detection of single photons. We assume that the modes $1$ and $2$ of a bipartite Gaussian state are superimposed to two additional modes, $3$ and $4$ respectively, at beam splitters of transmittivity $T$. For $T\rightarrow{1}$, the beam splitters tap at most a single photon from modes $1$ and $2$, this occurring with a small (but non-zero) chance.  The tapped photons populating modes 3 and 4 are then revealed at two single-photon resolving detectors. Upon measurement of one photon per mode, a single-photon subtracted state is achieved. By cascading $s$ of these stages, an $s$-photon subtracted state is conditionally produced and the operator $\hat{a}^s_1\hat{a}^s_2$ is implemented. While Eq.~(\ref{sottrattoformale}) gives a formal account of the effects of this operator onto a Gaussian state and, then, onto the entanglement transfer process, here we assess the fully physical case.  For the sake of simplicity, we address the single-photon case only. This is done by considering
\begin{equation}
\label{sottrattofisico}
\begin{aligned}
\rho^{''}_{12}&=\frac{{\cal N}}{\pi^2}\int{d}^2\xi{d}^2\eta~\chi(\xi,\eta){}_{3}\!\langle{1}|\hat{B}_{13}\hat{D}_1(-\xi)|0\rangle_{3}\langle{0}|\hat{B}_{13}^\dag|1\rangle_3\\
&\times{}_{4}\!\langle{1}|\hat{B}_{24}\hat{D}_2(-\eta)|0\rangle_{4}\langle{0}|\hat{B}^\dag_{24}|1\rangle_4,
\end{aligned}
\end{equation}
where we can write $\hat{D}_1(-\xi)|0\rangle_{3}\langle{0}|=\sum^\infty_{n,p=0}f_{np}(\xi)\proj{n,0}{13}{p,0}$ and $\hat{D}_2(-\eta)|0\rangle_{4}\langle{0}|=\sum^\infty_{m,q=0}f_{mq}(\eta)\proj{m,0}{24}{q,0}$. One can then use the Fock-state decomposition of the state resulting from the action of a beam splitter~\cite{barnettradmore}. For instance, we have that
\begin{equation}
\hat{B}_{13}|n,0\rangle_{13}\!=\!\sum^n_{k=0}(-1)^{n-k}T^{\frac{k}{2}}(1-T^2)^{\frac{n-k}{2}}\sqrt{\begin{pmatrix}n\\k\end{pmatrix}}\ket{k,n-k}_{13}.
\end{equation}
By putting everything together, we eventually get
\begin{equation}
\label{finalefisico}
\begin{aligned}
&\rho^{''}_{12}=\frac{(1-T^2)^2}{\pi^2}{\cal N}\!\sum^\infty_{n,p=0}\sum^\infty_{m,q=0}T^{\frac{n+m+p+q}{2}}\!\int{d}^2\xi{d}^2\eta\,\chi(\xi,\eta)\\
&\times\sqrt{(n+1)(m+1)(p+1)(q+1)}{f}_{n+1\,p+1}(\xi){f}_{m+1\,q+1}(\eta)\\
&\equiv{\frac{(1-T^2)^2}{\pi^2}{\cal N}\!\sum^\infty_{n,p=0}\sum^\infty_{m,q=0}T^{\frac{n+m+p+q}{2}}}\gamma^{p+1\,q+1}_{n+1\,m+1}.
\end{aligned}
\end{equation}
The factor $(1-T^2)^2$, which does not depend on the summations' indices, is washed out by the normalization factor ${\cal N}=\pi^2/[(1-T^2)^2\sum_{n,m}(n+1)(m+1)\gamma^{n+1\,m+1}_{n+1\,m+1}]$. From this point on, the time dependence of the two-qubit density matrix elements resulting from the entanglement transfer process can be evaluated as highlighted in the body of the manuscript. For $T\rightarrow{1}$ any of them becomes identical to the corresponding expression valid for the formal case, as studied in Sec.~\ref{nongaussiano}. Quantitatively, for $T\ge99.99\%$ the formal and physical approach give results that are indistinguishable from each other.



\begin{thebibliography}{99}

\bibitem{huelgacirac} J. I. Cirac, A. K. Ekert, S. F. Huelga, and C. Macchiavello, Phys. Rev. A {\bf 59}, 4249 (1999).

\bibitem{efforts} H. J. Kimble, Nature (London) {\bf 453}, 1023 (2008) and references within; R. J. Schoelkopf and S. M. Girvin, Nature (London) {\bf 451}, 664 (2008); H.-J. Briegel, W. D\"ur, J. I. Cirac, and P. Zoller, Phys. Rev. Lett. {\bf 81}, 5932 (1998); S. Olmschenk, D. N. Matsukevich, P. Maunz, D. Hayes, L.-M. Duan, and C. Monroe, Science {\bf 323}, 486 (2009).

\bibitem{ET} W. Son, J. Lee, M. S. Kim, D. Ahn, J. Mod. Opt. {\bf 49}, 1739 (2002); M. Paternostro, G. Falci, M. Kim, and G. M. Palma, Phys. Rev. B {\bf 69}, 214502 (2004); M. Paternostro, W. Son, M. S. Kim, G. Falci, and G. M. Palma, Phys. Rev. A {\bf 70}, 022320 (2004).

\bibitem{altri1} F. Casagrande, A. Lulli, and M. G. A. Paris, Phys. Rev. A {\bf 75}, 032336 (2007); Phys. Rev. A. {\bf 79}, 022307 (2009); M. Bina, F. Casagrande, M. Genoni, A. Lulli, and M. G. A. Paris, arXiv:0904.4317 (2009).

\bibitem{altri2} D. McHugh, M. Ziman, and V. Bu\v{z}ek, Phys. Rev. A {\bf 74}, 042303 (2006); D. Cavalcanti, J. G. Oliveira, J. G. de Faria, M. O. Cunha, and M. Fran\c{c}a Santos, Phys. Rev. A {\bf 74}, 042328 (2006).

\bibitem{kimble} K. S. Choi, H. Deng, J. Laurat, and H. J. Kimble, Nature {\bf 452}, 67 (2008).

\bibitem{kitagawa} A. Kitagawa, M. Takeoka, M. Sasaki, and A. Chefles, Phys.
Rev. A {\bf 73}, 042310 (2006).

\bibitem{furusawa} H. Takahashi, J. S. Neergaard-Nielsen, M. Takeuchi, M. Takeoka, K. Hayasaka, A. Furusawa, and M. Sasaki, arXiv:0907.2159v1 (2009).

\bibitem{paris} M. G. Genoni, M. G. A. Paris, and K. Banaszek, Phys. Rev. A {\bf 78}, 060303(R) (2008).

\bibitem{reciproco} J. Lee, M. Paternostro, S. Bose, and M. S. Kim, Phys. Rev. Lett. {\bf  96}, 080501 (2006).

\bibitem{barnettradmore} S. M. Barnett and P. M. Radmore, {\it Methods in Theoretical Quantum Optics} (Oxford University Press, New York, 1997).

\bibitem{glauber} K. E. Cahill and R. J. Glauber, Phys. Rev. {\bf 177}, 1857 (1969).

\bibitem{duan} L.-M. Duan, G. Giedke, J. I. Cirac, and P. Zoller, Phys. Rev. Lett. {\bf 84}, 2722 (2000).

\bibitem{adessoilluminati} G. Adesso and F. Illuminati, J. Phys. A {\bf 40}, 7821 (2007).

\bibitem{NPT} A. Peres, Phys. Rev. Lett. {\bf 77} 1413 (1996); M. Horodecki, P. Horodecki, and R. Horodecki, Phys. Lett. A {\bf 223}, 1 (1996); R. Simon, Phys. Rev. Lett. {\bf 84}, 2726 (2000).

\bibitem{schuck} E. Shchukin and W. Vogel, Phys. Rev. Lett. {\bf 95}, 230502 (2005).

\bibitem{serafini} A. Serafini, F. Illuminati, and S. De Siena, J. Phys. A {\bf 37}, L21 (2004).

\bibitem{oliveira} F. A. de Oliveira, M. S. Kim, P. L. Knight, and V. Bu\v{z}ek, Phys. Rev. A {\bf 41}, 1645 (1990).

\bibitem{JC} See B. W. Shore and P. L. Knight, J. Mod. Opt. {\bf 40}, 1195 (1993) for an excellent review and J. M. Fink, M. G\"oppl, M. Baur, R. Bianchetti, P. J. Leek, A. Blais, and A. Wallraff, Nature {\bf 454}, 315 (2008) for a very recent related experiment in the context of circuit quantum electrodynamics. 

\bibitem{phoenix} S. J. D. Phoenix and P. L. Knight, Ann. Phys. {\bf 186}, 381 (1988).

\bibitem{AS} M. Abramowitz and I. Stegun, {\it Handbook of Mathematical Functions with Formulas, Graphs, and Mathematical Tables} (Dover, New York, 1964)

\bibitem{GR} I. S. Gradshteyn and I. M. Ryzhik, {\it Tables of Integrals, Series, and Products} (Academic Press, New York, 1965).

\bibitem{measure} J. Lee, M. S. Kim, Y. J. Park, and S. Lee, J. Mod. Opt. {\bf 47}, 2151 (2000).

\bibitem{noi} S. Campbell, G. Adesso, and M. Paternostro (in preparation, 2009).

\bibitem{wallsmilburn} D. F. Walls and G. J. Milburn, {\it Quantum Optics}, (Springer Verlag, Heidelberg, 1995).

\bibitem{nota} Clearly, here our assumption is that a dissipation-affected CV resource is used in order to implement an otherwise dissipationless entanglement trnasfer process. The study of losses affecting the transfer mechanism are studied in~\cite{noi}.

\bibitem{serale} M. G. A. Paris, F. Illuminati, A. Serafini, and S. De Siena, Phys. Rev. A {\bf 68}, 012314 (2003); A. Serafini, F. Illuminati, M. G. A. Paris, and S. De Siena, Phys. Rev. A {\bf 69}, 022318 (2004); A. Serafini, M. G. A. Paris, F. Illuminati, and S. De Siena, J. Opt. B: Quantum Semiclass. Opt. {\bf 7}, R19 (2005).

\bibitem{refs}  T. Opatrny, G. Kurizki, and D.-G. Welsch, Phys. Rev. A {\bf 61}, 032302 (2000); P. T. Cochrane, T. C. Ralph, and G. J. Milburn, Phys. Rev. A {\bf 65}, 062306 (2002); S. Olivares, M. G. A. Paris, and R. Bonifacio, Phys. Rev. A {\bf 67}, 032314 (2003).

\bibitem{wolf} M. M. Wolf, G. Giedke, and J. I. Cirac, Phys. Rev. Lett. {\bf 96}, 080502 (2006).

\bibitem{adesso} G. Adesso, Phys. Rev. A {\bf 79}, 022315 (2009).

\bibitem{reciproco2} J. Lee, M. Paternostro, C. Ogden, Y. W. Cheong, S. Bose, and M. S. Kim, New J. Phys. {\bf 8}, 23 (2006); L. Zhou and G.-H. Yang, J. Phys. B: At. Mol. Opt. Phys. {\bf 39}, 5143 (2006); D. Ballester, Phys. Rev. A {\bf 79}, 062317 (2009).

\bibitem{tufarelli} T. Tufarelli, S. Bose, and M. S. Kim,  arXiv:0907.1831v1 (2009).

\end{thebibliography}
\end{document}